\documentclass[aps,prc,twocolumn,superscriptaddress,showpacs,floatfix,nofootinbib]{revtex4-1}
\usepackage{amsmath,graphicx,color,ulem}
\newcommand{\trento}{T$\mathrel{\protect\raisebox{-2.1pt}{R}}$ENTo}

\begin{document}

\title{Thermodynamics of hot strong-interaction matter \\ from ultrarelativistic nuclear collisions}

\author{Fernando G. Gardim}
\affiliation{Instituto de Ci\^encia e Tecnologia, Universidade Federal de Alfenas, 37715-400 Po\c cos de Caldas, MG, Brazil}
\affiliation{Institut de physique th\'eorique, Universit\'e Paris Saclay, CNRS, CEA, F-91191
Gif-sur-Yvette, France}
\author{Giuliano Giacalone}
\affiliation{Institut de physique th\'eorique, Universit\'e Paris Saclay, CNRS, CEA, F-91191
Gif-sur-Yvette, France}
\author{Matthew Luzum}
\affiliation{Instituto de F\'{i}sica, Universidade de S\~{a}o Paulo, 
R. do Mat\~{a}o 1371, 05508-090  S\~{a}o Paulo, SP, Brazil}
\author{Jean-Yves Ollitrault}
\affiliation{Institut de physique th\'eorique, Universit\'e Paris Saclay, CNRS, CEA, F-91191
Gif-sur-Yvette, France}
\begin{abstract}
  Collisions between heavy atomic nuclei at ultra-relativistic energies are carried out at particle colliders to produce the quark-gluon plasma, a state of matter where quarks and gluons are not confined into hadrons, and colour degrees of freedom are liberated.
  This state is thought to be produced as a transient phenomenon before it fragments into thousands of particles that reach the particle detectors. Despite two decades of investigations, one of the big open questions~\cite{Busza:2018rrf} is to obtain an experimental determination of the temperature reached in a heavy-ion collision, and a simultaneous determination of another thermodynamic quantity, such as the entropy density, that would give access to the number of degrees of freedom. Here we obtain the first such determination, utilizing state-of-the-art hydrodynamic simulations~\cite{Romatschke:2017ejr}. We define an effective temperature, averaged over the space-time evolution of the medium. Then, using experimental data, we determine this temperature, the corresponding entropy
density and speed of sound in the matter created in lead-lead collisions at the Large Hadron Collider. Our results agree with first-principles calculations from lattice quantum chromodynamics~\cite{Borsanyi:2013bia} and confirm that a deconfined phase of matter is indeed produced.
\end{abstract}

\maketitle
Relativistic hydrodynamics successfully explains the bulk of particle production in heavy-ion collisions. 
 The little droplet of quark-gluon plasma formed in a collision undergoes a very quick phase of thermalization~\cite{Schlichting:2019abc}  before expanding according to relativistic hydrodynamic equations~\cite{Romatschke:2017ejr}.
The fluid eventually decouples into individual particles as its density decreases~\cite{Broniowski:2001we}.
This freeze-out yields a number of resonances which decay into stable hadrons~\cite{Alba:2017hhe,Mazeliauskas:2018irt}.

It has been speculated for decades~\cite{VanHove:1982vk,Campanini:2011bj} that the temperature of the medium in the hydrodynamic phase is related to the mean momentum of the produced particles, $\langle p_t \rangle$, in the \textit{transverse} plane, i.e., the plane orthogonal to the collision axis. 
However, despite early attempts~\cite{McLerran:1986nc,Blaizot:1987cc}, this relation has never been explored in the context of hydrodynamic simulations.
Here, we identify this correspondence precisely. 
\begin{figure}
    \centering
    \includegraphics[width=\linewidth]{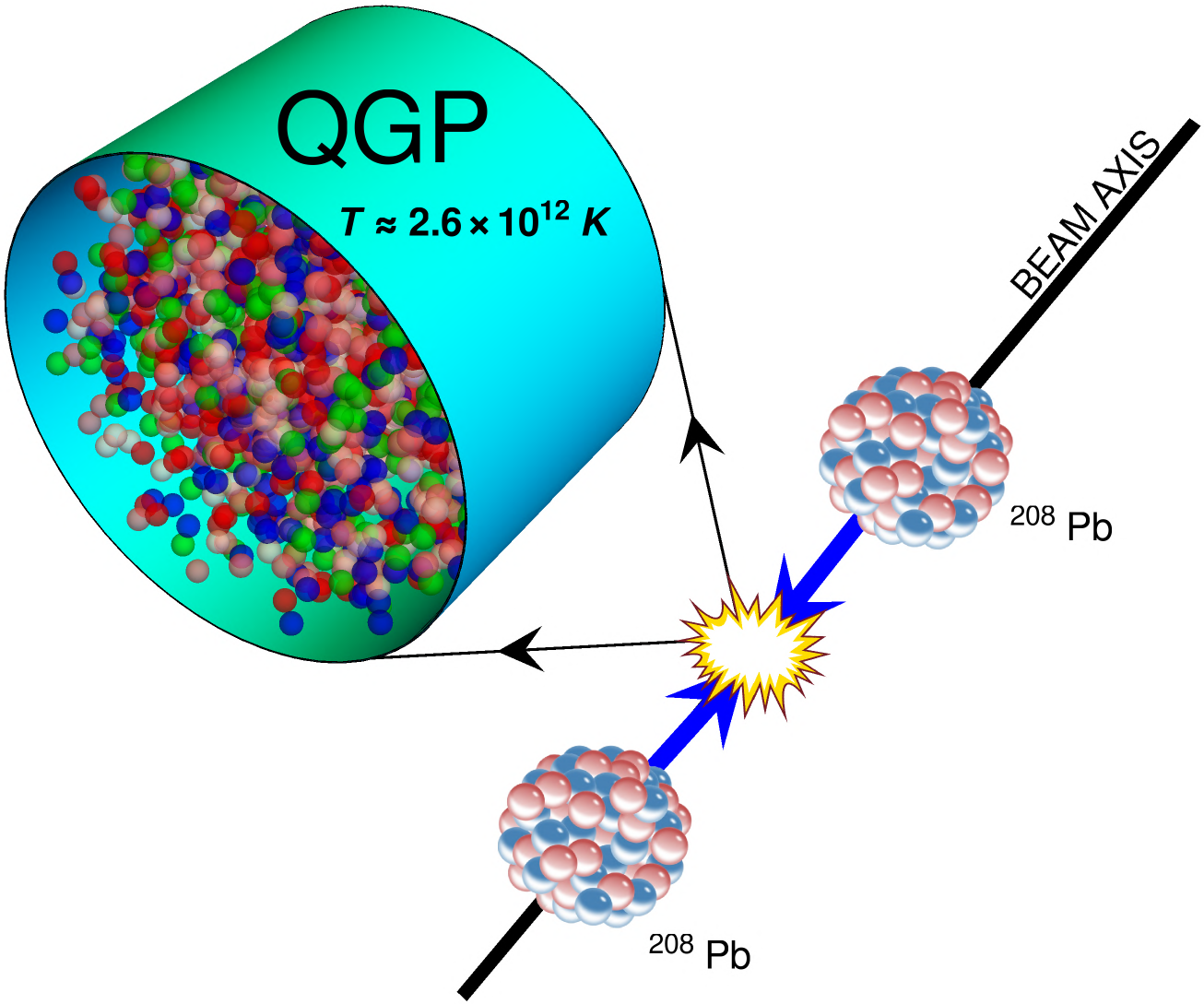}
    \caption{Schematic representation of a $^{208}$Pb-$^{208}$Pb collision at the Large Hadron Collider (LHC).
      The great density achieved in the collision produces a quark-gluon plasma (QGP), reaching temperatures of a few trillion Kelvin.
      We also display the effective volume, defined as a cylinder containing a uniform QGP at rest that would have the same energy and entropy as that produced in the collision.
      The coloured spheres are pictorial representations of the elementary particles (quarks and gluons) forming the QGP. 
      }
    \label{fig:fig1}
\end{figure}

In order to understand the physical picture, consider first the following thought experiment:
A uniform gas at rest is contained in a volume $V$ placed in the vacuum. 
At time $t=0$, the gas is allowed to expand freely.
If interactions are strong enough, this expansion is ruled by ideal (inviscid) hydrodynamics, and
both the total entropy and the total energy are conserved.
If one measures the total number of particles and the average energy per particle in the final state, one can reconstruct the initial energy density and entropy density, provided that one knows the initial volume $V$ and the entropy per particle.
Thus, one reconstructs the thermodynamics of the initial state from conservation laws, irrespective of details of the hydrodynamic expansion.

The situation in a nucleus-nucleus collision is slightly different:
The quark-gluon plasma is not produced at rest, but with a built-in longitudinal expansion imprinted by the collision dynamics, and detectors cover a limited angular range, so that only a slice of the fluid is seen.
The energy of this slice is not conserved, but decreases due to the negative work of pressure forces during the longitudinal expansion. 
As we shall see, however, the above argument still holds, provided that one replaces the initial energy with the final energy.

We thus evaluate the thermodynamics from the total energy, $E$, and total entropy, $S$, at freeze-out.
Precisely, we define the {\it effective\/} temperature, $T_{\rm eff}$, and the effective volume, $V_{\rm eff}$, as those of a uniform fluid at rest which would have the same energy and entropy as the fluid at freeze-out (see the illustration in Fig.~\ref{fig:fig1}).
They are defined by the equations
\begin{eqnarray}
  \label{effective}
E=\int_{\rm f.o.} T^{0\mu} d\sigma_\mu&=&\epsilon(T_{\rm eff}) V_{\rm eff},\cr
S=\int_{\rm f.o.} s u^{\mu} d\sigma_\mu&=&s(T_{\rm eff}) V_{\rm eff},
\end{eqnarray}
where the integrals run over the freeze-out hypersurface.
$T^{\mu\nu}$ denotes the stress-energy tensor of the fluid and $u^{\mu}$ the fluid 4-velocity~\cite{Ollitrault:2008zz}. 
$\epsilon$ and $s$ denote, respectively, the energy and entropy density in the fluid rest frame. 
By taking the ratio $E/S$, one eliminates $V_{\rm eff}$, and one can solve the resulting equation for $T_{\rm eff}$, using the same equation of state as in the hydrodynamic calculation.
Note that $T_{\rm eff}$ and $s(T_{\rm eff})$ are related by the equation of state of the fluid {\it by construction.\/}
The effective temperature is smaller than the initial temperature because of the longitudinal cooling.
On the other hand, it is larger than the freeze-out temperature, because the energy $E$ defined by Eq.~(\ref{effective}) contains the kinetic energy due to the collective motion of the fluid. 
\begin{figure*}
    \centering
    \includegraphics[width=\linewidth]{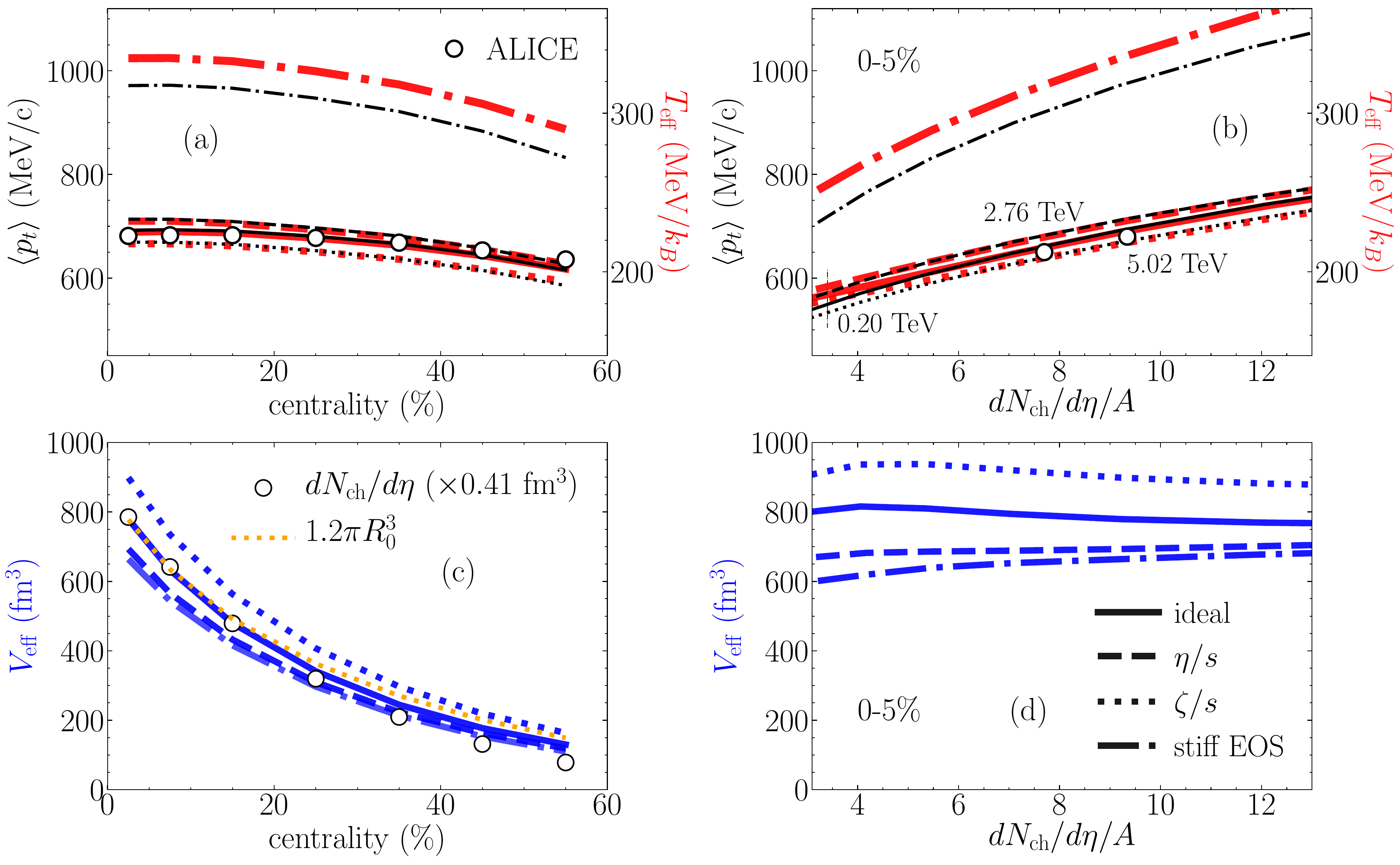}
    \caption{Results from hydrodynamic simulations of Pb+Pb collisions. Black curves correspond to the average transverse momentum, $\langle p_t\rangle$, red curves to the effective temperature, $T_{\rm eff}$, and blue curves to the effective volume, $V_{\rm eff}$.
      In panels (a) and (c), results are presented as a function of the collision centrality for a fixed energy per nucleon-nucleon collision~$\sqrt{s_{\rm NN}}=5.02$~TeV, corresponding to the current energy available at the LHC for ion beams.
      In panels (b) and (d), results are presented as a function of the collision energy for a fixed centrality range (5\% most central collisions). 
The variable on the horizontal axis is not the collision energy itself, but the multiplicity of charged particles in the pseudorapidity interval $-0.5<\eta<0.5$, or, equivalently (since $\eta\equiv-\ln\tan(\theta/2)$), in the range of polar angles $|\theta-90^\circ|<27.5^\circ$ relative to the beam.
This multiplicity is denoted by $dN_{\rm ch}/d\eta$.
We scale it by the mass number $A$ of the colliding nuclei, so that the top panels would be essentially identical for other symmetric collisions, such as Xe+Xe, Cu+Cu or Au+Au collisions~\cite{Rogly:2018lji,Giacalone:2019ldn}. 
      In panels (a) and (b), the scales for $\langle p_t\rangle$ and $T_{\rm eff}$ differ (see vertical scales to the left and right) by a constant factor $3.07$. Solid, dashed and dotted lines correspond to simulations using a realistic equation of state, representing respectively, ideal hydrodynamics, viscous hydrodynamics with shear viscosity $\eta$, and viscous hydrodynamics with bulk viscosity $\zeta$. $s$ denotes the entropy density. Dot-dashed lines correspond to ideal hydrodynamic simulations with the stiff equation of state. Symbols in the top panels are experimental LHC data for $\langle p_t\rangle$~\cite{Acharya:2018eaq}. The experimental value of $(1/A)dN_{\rm ch}/d\eta$ at centre-of-mass collision energy 200~GeV~\cite{Back:2002uc} is indicated as a vertical line in panel (b). Symbols in panel (c) represent (dimensionless) LHC data for the charged multiplicity, $dN_{\rm ch}/d\eta$~\cite{Adam:2015ptt}, multiplied by $0.41$~fm$^3$ (1~fm$=10^{-15}$~m), while the thin dotted line is $1.2\pi R_0^3$, where $R_0$ is defined by Eq.~(\ref{defR0}). Experimental errors are smaller than the symbol size.}
    \label{fig:fig2}
\end{figure*}

The energy per particle is $3T$ in a massless ideal gas at temperature $T$ with Boltzmann statistics~\cite{Ollitrault:2008zz}. 
We use natural units where $k_B = c = 1$, such that momentum and temperature have the same dimension.
The transverse momentum of a particle coincides with its energy in the ultrarelativistic limit and near midrapidity (i.e., for particles emitted perpendicular to the beam). Therefore, it is natural to expect that $\langle p_t\rangle\sim 3 T_{\rm eff}$.

To illustrate and verify this expectation, we carry out hydrodynamic simulations of Pb+Pb collisions. The details of the hydrodynamic setup are reported in the Methods.
Since the transport coefficients of the quark-gluon plasma are poorly constrained~\cite{Bernhard:2016tnd}, we carry out three different sets of calculations: ideal hydrodynamics, viscous hydrodynamics with shear viscosity only, and viscous hydrodynamics with bulk viscosity only. 
In our calculations we use mainly a soft equation of state expected from the theory of strong interactions~\cite{Huovinen:2009yb}, but we also perform a cross check of the results using a stiff equation of state where the speed of sound takes the maximum value $c_s=1/\sqrt{3}$, corresponding to a gas of massless particles. 

The results of our simulations are displayed in Fig.~\ref{fig:fig2}.
Panels (a) and (b) confirm the expectation that the effective temperature $T_{\rm eff}$ (red lines) is tightly correlated with the mean transverse momentum of charged particles, $\langle p_t\rangle$ (black lines). 
With the soft equation of state, black lines and red lines overlap, corresponding to the proportionality $\langle p_t\rangle=3.07~T_{\rm eff}$. 
Shear viscosity increases $\langle p_t\rangle$, while bulk viscosity decreases it~\cite{Monnai:2009ad}. 
The remarkable result is that $T_{\rm eff}$ is modified by the same relative amount, so that $\langle p_t\rangle=3.07~T_{\rm eff}$ holds irrespective of transport coefficients. 
This proportionality is satisfied for all centralities, and for most of the considered range of collision energies.
Deviations at the level of few percent appear only at Relativistic Heavy Ion Collider (RHIC) energies. 
The results using the stiff equation of state are displayed as dot-dashed lines. We note that, although $\langle p_t\rangle$ and $T_{\rm eff}$ increase by $\sim 40\%$, the proportionality coefficient only changes by $\sim 6\%$:
$\langle p_t\rangle\simeq 2.90~T_{\rm eff}$.

Our results for the effective volume, $V_{\rm eff}$, are displayed in panels (c) and (d) of Fig.~\ref{fig:fig2}.
This quantity is essentially determined by the initial radius, $R_0$, which we define by 
\begin{equation}
  \label{defR0}
(R_0)^2\equiv \frac{2\int_{\bf r} |{\bf r}|^2 s(\tau_0,{\bf r})}{\int_{\bf r} s(\tau_0,{\bf r})},
\end{equation}
where $s(\tau_0,{\bf r})$ is the entropy density at the time $\tau_0$ when the hydrodynamic evolution starts, and the integration runs over the transverse plane.
The factor $2$ in the numerator ensures that for a uniform density profile in a circle of radius $R_0$, the right-hand side gives $(R_0)^2$.
For dimensional reasons, the volume $V_{\rm eff}$ is proportional to $(R_0)^3$.
This explains the decrease of $V_{\rm eff}$ as a function of centrality percentile (this decrease follows that of the multiplicity, see Fig.~\ref{fig:fig2} (c)), and that $V_{\rm eff}$ is essentially independent of the collision energy (Fig.~\ref{fig:fig2} (d)).
The proportionality constant between $V_{\rm eff}$ and $R_0^3$ depends on details of the hydrodynamic modeling (transport coefficients and equation of state), although weakly.

The main conclusion from Fig.~\ref{fig:fig2} is that, while $\langle p_t\rangle$ and $T_{\rm eff}$ are sensitive to details of the hydrodynamic modeling (initial density profile, transport coefficients, equation of state), most of the model dependence disappears when considering the ratio $\langle p_t\rangle/T_{\rm eff}$.  
Using $T_{\rm eff}\simeq \langle p_t\rangle/3.07$ from Fig.~\ref{fig:fig2} (a) in combination with the experimental result $\langle p_t\rangle=681$~MeV~\cite{Acharya:2018eaq}, we obtain
\begin{eqnarray}
\label{eq:T}
T_{\rm eff}&=&222\pm 9~{\rm MeV}\cr
&=&(2.58\pm 0.10)\times 10^{12}~{\rm K} 
\end{eqnarray}
in central Pb+Pb collisions at $\sqrt{s_{\rm NN}}=5.02$~TeV.
The experimental observation that $\langle p_t\rangle$ is almost independent of centrality implies in turn that different centralities correspond to the same $T_{\rm eff}$. 
\begin{figure}
    \centering
    \includegraphics[width=\linewidth]{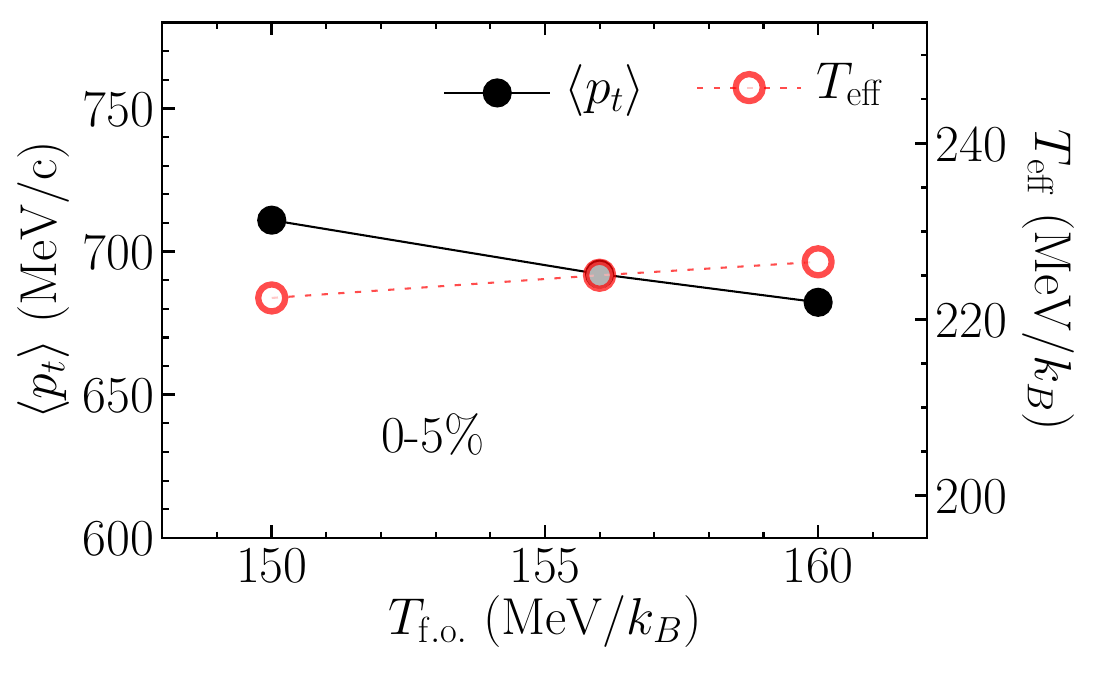}
    \caption{Estimate of the theoretical uncertainty on the effective temperature. 
Variation of $\langle p_t\rangle$ and $T_{\rm eff}$ as a function of the freeze-out temperature in ideal hydrodynamic simulations of central Pb+Pb collisions at $\sqrt{s_{\rm NN}}=5.02$~TeV.
Lines are drawn to guide the eye. 
    }
    \label{fig:tf}
\end{figure}

The uncertainty in Eq.~(\ref{eq:T}) is from the ratio $\langle p_t\rangle/T_{\rm eff}$.
This ratio depends slightly on the freeze-out temperature $T_{\rm f.o.}$ at which the fluid is transformed into individual particles.
Our default value is $T_{\rm f.o.}=156.5$~MeV (see Methods), but one must allow for a variation of $T_{\rm f.o.}$. 
Thermal fits to particle abundances return a temperature close to $160$~MeV~\cite{Andronic:2017pug}, which sets a natural upper limit on $T_{\rm f.o.}$. 
On the other hand, any value lower than $150$~MeV would miss the particle ratios after resonance decays by a significant amount. 
Figure~\ref{fig:tf} displays the variation of $T_{\rm eff}$ and $\langle p_t\rangle$ within this range. 
As explained above, $T_{\rm eff}$ is larger than  $T_{\rm f.o.}$ by construction. 
We observe that $T_{\rm eff}$ is almost {\it independent\/} of $T_{\rm f.o.}$, showing a mild increase in Fig.~\ref{fig:tf}, while $\langle p_t\rangle$ decreases.
We have checked that these results change very little if partial chemical equilibrium is implemented~\cite{Huovinen:2009yb}. 
We conclude that the error on the ratio $\langle p_t\rangle/T_{\rm eff}$ due to the freeze-out temperature is about 4\%, which sets the uncertainty in Eq.~(\ref{eq:T}).
Note that with a stiff equation of state, $\langle p_t\rangle/T_{\rm eff}$ is smaller by 6\%.
However, since the stiff equation of state overestimates $\langle p_t\rangle$ by 40\%, we do not include this 6\% difference in our uncertainty estimate in Eq.~(\ref{eq:T}).

Before evaluating quantitatively the equation of state, we perform a back-of-the-envelope estimate of the number of degrees of freedom. 
A quark-gluon plasma, modeled as a massless ideal gas with Boltzmann statistics, has a particle density $n=gT^3/\pi^2$~\cite{Ollitrault:2008zz}, where $g$ is the number of degrees of freedom (colour, flavour, spin), and we set $\hbar = 1$. 
If the number of produced hadrons is equal to the number of quarks and gluons in the medium, and by taking into account that only two thirds of the hadrons are charged, the particle density in the effective volume is $n=1.5(dN_{\rm ch}/d\eta)/V_{\rm eff}\sim 4~{\rm fm}^{-3}$, using $V_{\rm eff}\approx 780$~fm$^3$ from Fig.~\ref{fig:fig2}(c), and $dN_{\rm ch}/d\eta\approx 2000$ from experimental data. 
Now, with $T_{\rm eff}=222$~MeV and $\hbar c=197$~MeV$\cdot$fm, one obtains $g\sim30$.
This large number shows that the colour degrees of freedom are active, or, in other words, that a deconfined state is produced.

\begin{figure}
    \centering
    \includegraphics[width=.9\linewidth]{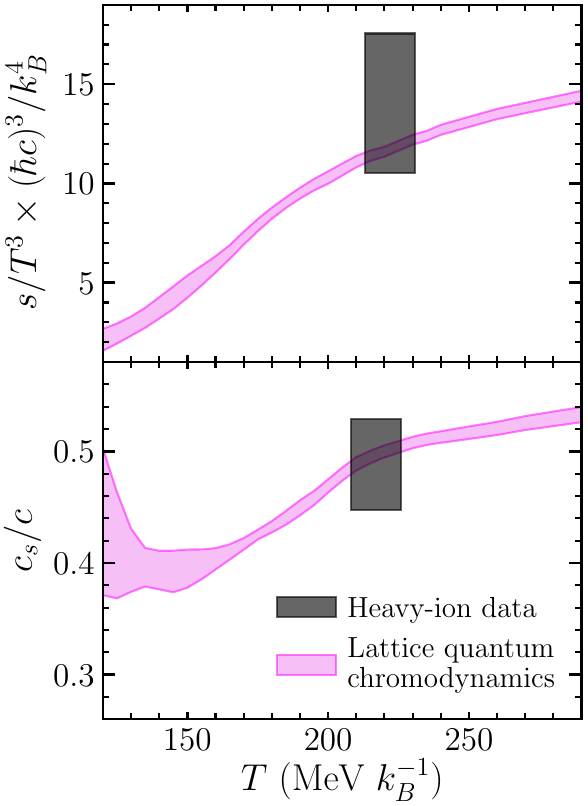}
    \caption{Thermodynamic properties of hot strong-interaction matter --- (top) entropy density scaled by $T^3$ and (bottom) speed of sound as a function of the temperature $T$.
      Magenta bands are ab-initio calculations using quantum chromodynamics (QCD) discretised on a lattice~\cite{Borsanyi:2013bia}, where the width of the band is the uncertainty. 
      Grey boxes display our results obtained from heavy-ion experimental data with the corresponding uncertainties (see text), as given by Eqs.~(\ref{eq:T}), (\ref{satteff}) and (\ref{cs}).
      Since the derivative in Eq.~(\ref{cs}) is evaluated using the evolution of observables from $\sqrt{s_{\rm NN}}=2.76$ to $\sqrt{s_{\rm NN}}=5.02$~TeV, the relevant temperature is half-way between these two points, i.e., 5~MeV lower than that given by Eq.~(\ref{eq:T}), and the box in the bottom panel has been shifted to the left accordingly.}
      \label{fig:lattice}
\end{figure}

We now evaluate the entropy density at $T_{\rm eff}$ from experimental data.
The charged multiplicity gives a direct measure of the entropy at freeze-out.
The effective entropy density is then related to the multiplicity through the formula:
\begin{equation}
  \label{seff}
s(T_{\rm eff})=\frac{1}{V_{\rm eff}}\frac{S}{N_{\rm ch}}\frac{dN_{\rm ch}}{dy},
\end{equation}
where $dN_{\rm ch}/dy$ denotes the multiplicity per unit rapidity (see Methods), while for the entropy per particle we use $S/N_{\rm ch}=6.7\pm 0.8$, obtained in a recent analysis~\cite{Hanus:2019fnc}. 
The theoretical uncertainty on $V_{\rm eff}$ depends on both transport coefficients and initial conditions. 
The uncertainty due to transport coefficients can be evaluated from Fig.~\ref{fig:fig2} (d) and is of the order of 15\%.
The uncertainty due to the initial size is comparable.
Different models of initial conditions give values of $R_0$ (see Eq.~(\ref{defR0})) that differ from the one used in our calculation by up to 3.5\%.
Since $V_{\rm eff}$ is proportional to $R_0^3$, this results in a 11\% uncertainty on $V_{\rm eff}$.
Taking $dN_{\rm ch}/d\eta$ from ALICE data~\cite{Adam:2015ptt} and $V_{\rm eff}$ from the ideal hydrodynamic calculation in Fig.~\ref{fig:fig2} (d), assuming that $dN/dy\simeq 1.15 dN/d\eta$ near midrapidity~\cite{Hanus:2019fnc}, and adding all errors in quadrature, we obtain
\begin{equation}
  \label{satteff}
  s(T_{\rm eff})=20\pm 5~{\rm fm}^{-3}.
\end{equation}
With Eq.~(\ref{eq:T}), this gives $s(T_{\rm eff})/T_{\rm eff}^3=14\pm 3.5$,
in agreement with ab-initio calculations (see Fig.~\ref{fig:lattice}, top).

We finally evaluate the speed of sound by varying the collision energy. 
Increasing energy at a fixed centrality amounts to stuffing more energy in a fixed volume, and the associated increase in pressure or temperature gives the compressibility, which is related to the speed of sound.
Quite remarkably, all large sources of uncertainty (freeze-out temperature, transport coefficients, system size) cancel if one considers the relative variation of $\langle p_t\rangle$ and $dN_{\rm ch}/d\eta$ with colliding energy, rather than their absolute values.
The effective volume is essentially constant as a function of energy, so that the effective entropy density is proportional to the multiplicity; Analogously, the effective temperature is proportional to $\langle p_t\rangle$. This implies:
\begin{equation}
  \label{dseff}
\frac{ds(T_{\rm eff})}{s(T_{\rm eff})}=\frac{dN_{\rm ch}}{N_{\rm ch}},\hspace{40pt}\frac{dT_{\rm eff}}{T_{\rm eff}}=\frac{d\langle p_t\rangle}{\langle p_t\rangle}.
\end{equation}
These relative variations do not involve the proportionality coefficients, which are the dominant source of error for both quantities. 
Taking the ratio of the right-hand sides in Eq.~(\ref{dseff}), one obtains the velocity of sound at $T_{\rm eff}$~\cite{Campanini:2011bj}:
\begin{equation}
  \label{cs2}
  c_s^2(T_{\rm eff})\equiv \frac{dP}{d\varepsilon}=
  \left.\frac{sdT}{Tds}\right|_{T_{\rm eff}}=\frac{d\ln\langle p_t\rangle}{d\ln (dN_{\rm ch}/d\eta)}.
\end{equation}
We estimate the right-hand side using 0-5\% central Pb-Pb data.
We use the values of $dN_{\rm ch}/d\eta$
at 2.76~TeV~\cite{Aamodt:2010cz} and 5.02~TeV~\cite{Adam:2015ptt}, which are extrapolated down to $p_t=0$, and $p_t$ spectra from Ref.~\cite{Acharya:2018qsh}, without carrying any extrapolation down to $p_t=0$, i.e., assuming that the relative variation of $\langle p_t\rangle$ is insensitive to the extrapolation.
We obtain
\begin{equation}
  \label{cs}
  c_s^2(T_{\rm eff})=0.24\pm 0.04,
\end{equation}
where the error bar comes from taking into account the possible variation of the effective volume between the two energies, which we estimate to be at most 3\% according to the results in Fig.~\ref{fig:fig2} (d).
Thus, the speed of sound in the quark-gluon plasma produced at the Large Hadron Collider (LHC) is half the speed of light in vacuum, in agreement with ab-initio calculations (see Fig.~\ref{fig:lattice}, bottom).

Note that Eq.~(\ref{cs2}) elucidates the conclusion from the Bayesian analysis of Ref.~\cite{Pratt:2015zsa} that RHIC and LHC data in combination would provide a better constraint on the speed of sound than either alone.
We have not analysed RHIC data for the following reason: The latest data on spectra~\cite{Adare:2013esx} are limited to $p_t>0.5$~GeV, and the extrapolation down to $p_t=0$ would entail large errors.
We emphasise that if measurements of $\langle p_t\rangle$ were available at RHIC,
it would be straightforward---and potentially of
great interest---to extend our analysis down to the lowest
beam energies, to probe QCD thermodynamics at other
temperatures and at nonzero baryon chemical potential.

\section*{Acknowledgments}
F.G.G. was supported by Conselho Nacional de Desenvolvimento Cient\'{\i}fico  e  Tecnol\'ogico  (CNPq grant 205369/2018-9 and 312932/2018-9). 
M.L.~acknowledges support from FAPESP projects 2016/24029-6  and 2017/05685-2.
F.G.G. and  M.L.  acknowledge support from project INCT-FNA Proc.~No.~464898/2014-5 and
G.G., M.L. and J.-Y.O.   from  USP-COFECUB (grant Uc Ph 160-16, 2015/13).



\section*{Methods}
The details of the hydrodynamic calculation are as follows. 
We assume a boost-invariant longitudinal expansion~\cite{Bjorken:1982qr}, and we neglect the baryon chemical potential.
These are good approximations at ultrarelativistic energies.
Longitudinal boost invariance implies that the quantities $E$, $S$, $V_{\rm eff}$ are per unit rapidity, i.e., they stand for $dE/dy$, $dS/dy$, $dV_{\rm eff}/dy$ (see also Eq.~(\ref{seff})). 
We initialise hydrodynamics at a time $\tau_0=0.6$~fm/$c$~\cite{Kolb:2003dz} after the collision. 
We neglect the transverse expansion before $\tau_0$~\cite{Vredevoogd:2008id,vanderSchee:2013pia,Keegan:2016cpi}.
We assume that the entropy density at time $\tau_0$ at a given transverse point $\bf{r}$, $s(\tau_0,\bf r)$, is proportional to $\sqrt{T_AT_B}$ computed at $\bf{r}$~\cite{Eskola:2000xq,Kolb:2001qz}, where $A$ and $B$ label the two colliding nuclei, and $T_{A/B}$ is the integral of the nuclear density along the longitudinal coordinate (or, equivalently, the thickness function in the optical Glauber model~\cite{Miller:2007ri}).
The proportionality factor is adjusted at each collision centrality so as to match the observed charged multiplicity $dN_{\rm ch}/d\eta$ in the pseudorapidity interval $|\eta|<0.5$~\cite{Adam:2015ptt,Acharya:2018hhy}. 

The centrality is specified by the impact parameter, $b$, of the collision. The relation between centrality fraction, $c$, and the impact parameter is geometric, $c=\pi b^2/\sigma_{\rm PbPb}$, $\sigma_{\rm PbPb}=767$~fm$^2$ being the total inelastic cross section of Pb-Pb collisions at $\sqrt{s_{\rm NN}}=5.02$~TeV~\cite{Abelev:2013qoq}.
We run a single hydrodynamic event at each centrality.

The choice of taking $\sqrt{T_AT_B}$ is motivated by the phenomenological success of the \trento{} model of initial conditions \cite{Moreland:2014oya}, where the same prescription for entropy deposition is used, although with the inclusion of initial-state fluctuations~\cite{Hama:2004rr}.
We include initial-state fluctuations in our smooth calculation through their effect on the transverse size of the system. 
The mean transverse momentum in hydrodynamics is sensitive to the transverse size~\cite{Broniowski:2009fm}, which is somewhat reduced by initial state fluctuations.
To take into account this reduction, at each impact parameter we shrink our smooth initial conditions so that they present the same radius $R_0$ (see below Eq.~(\ref{defR0})) as in the full \trento{} parametrization, tuned to data as in Ref.~\cite{Giacalone:2017dud}.
The correction to the size is of order 5\%, 15\%, 30\% at $b=2$, $7$, $12$~fm, respectively.

The fluctuation-corrected initial conditions are then evolved using the MUSIC hydrodynamic code~\cite{Schenke:2010nt,Schenke:2011bn,Paquet:2015lta}.
We run it either with a realistic soft equation of state computed using lattice QCD, s95p-v1~\cite{Huovinen:2009yb}, or with a stiff equation of state:
\begin{equation}
  \label{stiff}
\epsilon=3P+C,
\end{equation}
where $C$ is a constant adjusted so that $P$ and $\epsilon$ at freeze-out are correct for a gas of hadrons and resonances (as with the soft equation of state), so that energy and momentum are conserved. 

We run ideal and viscous hydrodynamic simulations.
In order to assess separately the effects of shear and bulk viscosity, we implement either a constant shear viscosity over entropy ratio, $\eta/s=0.2$~\cite{Heinz:2013th,Niemi:2015qia}, or a bulk viscosity parametrised as in Ref.~\cite{Bernhard:2016tnd}. 
Cooper-Frye freeze-out~\cite{Cooper:1974mv} is implemented at the temperature $T_{\rm f.o.}=156.5$~MeV~\cite{Bazavov:2018mes}, a reasonable choice in a hydrodynamic setup that does not implement partial chemical equilibrium~\cite{Huovinen:2009yb,Gale:2012rq,Heinz:2013th,Eskola:2017bup} or a hadronic cascade~\cite{Weller:2017tsr,Dubla:2018czx}.
The viscous corrections to the momentum distribution functions are evaluated using the quadratic ansatz ~\cite{Teaney:2003kp,Dusling:2009df,Bozek:2009dw}.
We take into account hadronic decays after freeze-out, but we neglect rescatterings in the hadronic phase~\cite{Teaney:2000cw,Petersen:2008dd,Bernhard:2016tnd}.

We evaluate the mean transverse momentum of charged hadrons in the pseudorapidity interval $|\eta|<0.8$. 
This observable does not require particle identification, unlike the mean transverse mass~\cite{Petersen:2009mz,Luo:2015doi,Monnai:2017cbv}, but it requires to cover the whole $p_t$ range, and the low $p_t$ particles are not detected.
There are two strategies in order to compare theory and experiment: 
One can either implement the same $p_t$ cut in the calculation, or extrapolate experimental data down to $p_t=0$.
We choose the latter approach, because hydrodynamics is meant to describe the bulk of particle production, and because this extrapolation has been carried out by some of the experiments.



\section*{Code Availability}

The hydrodynamic code used in this article is publicly available at the following address:
{\tt https://sourceforge.net/projects/music-hydro/}


\begin{thebibliography}{9}


\bibitem{Busza:2018rrf} 
  W.~Busza, K.~Rajagopal and W.~van der Schee,
  ``Heavy Ion Collisions: The Big Picture, and the Big Questions,''
  Ann.\ Rev.\ Nucl.\ Part.\ Sci.\  {\bf 68}, 339-376 (2018).



\bibitem{Romatschke:2017ejr} 
  P.~Romatschke \& U.~Romatschke,
  {\it Relativistic Fluid Dynamics In and Out of Equilibrium}
  (Cambridge University Press, 2019).



\bibitem{Borsanyi:2013bia} 
  S.~Borsanyi, Z.~Fodor, C.~Hoelbling, S.~D.~Katz, S.~Krieg and K.~K.~Szabo,
  ``Full result for the QCD equation of state with 2+1 flavors,''
  Phys.\ Lett.\ B {\bf 730}, 99-104 (2014).


\bibitem{Schlichting:2019abc} 
  S.~Schlichting and D.~Teaney,
  ``The First fm/c of Heavy-Ion Collisions,''
  Ann.\ Rev.\ Nucl.\ Part.\ Sci.\  {\bf 69}, 447-476 (2019)



\bibitem{Broniowski:2001we} 
  W.~Broniowski and W.~Florkowski,
  ``Explanation of the RHIC p(T) spectra in a thermal model with expansion,''
  Phys.\ Rev.\ Lett.\  {\bf 87}, 272302 (2001).



\bibitem{Alba:2017hhe} 
  P.~Alba, V.~Mantovani Sarti, J.~Noronha, J.~Noronha-Hostler, P.~Parotto, I.~Portillo Vazquez and C.~Ratti,
  ``Effect of the QCD equation of state and strange hadronic resonances on multiparticle correlations in heavy ion collisions,''
  Phys.\ Rev.\ C {\bf 98}, no. 3, 034909 (2018).



\bibitem{Mazeliauskas:2018irt} 
  A.~Mazeliauskas, S.~Floerchinger, E.~Grossi and D.~Teaney,
  ``Fast resonance decays in nuclear collisions,''
  Eur.\ Phys.\ J.\ C {\bf 79}, no. 3, 284 (2019).



\bibitem{VanHove:1982vk} 
  L.~Van Hove,
  ``Multiplicity Dependence of p(T) Spectrum as a Possible Signal for a Phase Transition in Hadronic Collisions,''
  Phys.\ Lett.\ B {\bf 118}, 138-140 (1982).



\bibitem{Campanini:2011bj} 
  R.~Campanini and G.~Ferri,
  ``Experimental equation of state in proton-proton and proton-antiproton collisions and phase transition to quark gluon plasma,''
  Phys.\ Lett.\ B {\bf 703}, 237-245 (2011).



\bibitem{McLerran:1986nc} 
  L.~D.~McLerran, M.~Kataja, P.~V.~Ruuskanen and H.~von Gersdorff,
  ``Studies of the Hydrodynamical Evolution of Matter Produced in Fluctuations in p anti-p Collisions and in Ultrarelativistic Nuclear Collisions. 2. Transverse Momentum Distributions,''
  Phys.\ Rev.\ D {\bf 34}, 2755-2763 (1986).



\bibitem{Blaizot:1987cc} 
  J.~P.~Blaizot and J.~Y.~Ollitrault,
  ``Equation of State and Hydrodynamics of Quark Gluon Plasmas,''
  Phys.\ Lett.\ B {\bf 191}, 21-26 (1987).



\bibitem{Ollitrault:2008zz} 
  J.~Y.~Ollitrault,
  ``Relativistic hydrodynamics for heavy-ion collisions,''
  Eur.\ J.\ Phys.\  {\bf 29}, 275-302 (2008).



\bibitem{Acharya:2018eaq} 
  S.~Acharya {\it et al.} [ALICE Collaboration],
  ``Transverse momentum spectra and nuclear modification factors of charged particles in Xe-Xe collisions at $\sqrt{s_{\rm NN}}$ = 5.44 TeV,''
  Phys.\ Lett.\ B {\bf 788}, 166-179 (2019)



\bibitem{Back:2002uc} 
  B.~B.~Back {\it et al.} [PHOBOS Collaboration],
  ``Centrality dependence of the charged particle multiplicity near mid-rapidity in Au + Au collisions at $\sqrt{s}$ (NN) = 130-GeV and 200-GeV,''
  Phys.\ Rev.\ C {\bf 65}, 061901 (2002)



\bibitem{Adam:2015ptt} 
  J.~Adam {\it et al.} [ALICE Collaboration],
  ``Centrality dependence of the charged-particle multiplicity density at midrapidity in Pb-Pb collisions at $\sqrt{s_{\rm NN}}$ = 5.02 TeV,''
  Phys.\ Rev.\ Lett.\  {\bf 116}, no. 22, 222302 (2016)



\bibitem{Bernhard:2016tnd} 
  J.~E.~Bernhard, J.~S.~Moreland, S.~A.~Bass, J.~Liu and U.~Heinz,
  ``Applying Bayesian parameter estimation to relativistic heavy-ion collisions: simultaneous characterization of the initial state and quark-gluon plasma medium,''
  Phys.\ Rev.\ C {\bf 94}, no. 2, 024907 (2016)

\bibitem{Huovinen:2009yb} 
  P.~Huovinen and P.~Petreczky,
  ``QCD Equation of State and Hadron Resonance Gas,''
  Nucl.\ Phys.\ A {\bf 837}, 26-53 (2010)

\bibitem{Monnai:2009ad} 
  A.~Monnai and T.~Hirano,
  ``Effects of Bulk Viscosity at Freezeout,''
  Phys.\ Rev.\ C {\bf 80}, 054906 (2009)



\bibitem{Niemi:2015qia} 
  H.~Niemi, K.~J.~Eskola and R.~Paatelainen,
  ``Event-by-event fluctuations in a perturbative QCD + saturation + hydrodynamics model: Determining QCD matter shear viscosity in ultrarelativistic heavy-ion collisions,''
  Phys.\ Rev.\ C {\bf 93}, no. 2, 024907 (2016)



\bibitem{Acharya:2018hhy} 
  S.~Acharya {\it et al.} [ALICE Collaboration],
  ``Centrality and pseudorapidity dependence of the charged-particle multiplicity density in Xe-Xe collisions at $\sqrt{s_{\rm NN}}$ =5.44TeV,''
  Phys.\ Lett.\ B {\bf 790}, 35-48 (2019)



\bibitem{Giacalone:2017dud} 
  G.~Giacalone, J.~Noronha-Hostler, M.~Luzum and J.~Y.~Ollitrault,
  ``Hydrodynamic predictions for 5.44 TeV Xe+Xe collisions,''
  Phys.\ Rev.\ C {\bf 97}, no. 3, 034904 (2018)



\bibitem{Rogly:2018lji} 
  R.~Rogly, G.~Giacalone and J.~Y.~Ollitrault,
  ``Geometric scaling in symmetric nucleus-nucleus collisions,''
  Nucl.\ Phys.\ A {\bf 982}, 355-358 (2019)



\bibitem{Giacalone:2019ldn} 
  G.~Giacalone, A.~Mazeliauskas and S.~Schlichting,
  ``Hydrodynamic attractors, initial state energy and particle production in relativistic nuclear collisions,''
  Phys.\ Rev.\ Lett.\  {\bf 123}, no. 26, 262301 (2019)



\bibitem{Andronic:2017pug} 
  A.~Andronic, P.~Braun-Munzinger, K.~Redlich and J.~Stachel,
  ``Decoding the phase structure of QCD via particle production at high energy,''
  Nature {\bf 561}, no. 7723, 321-330 (2018).



\bibitem{Hanus:2019fnc} 
  P.~Hanus, A.~Mazeliauskas and K.~Reygers,
  ``Entropy production in pp and Pb-Pb collisions at energies available at the CERN Large Hadron Collider,''
  Phys.\ Rev.\ C {\bf 100}, no. 6, 064903 (2019)



\bibitem{Aamodt:2010cz} 
  K.~Aamodt {\it et al.} [ALICE Collaboration],
  ``Centrality dependence of the charged-particle multiplicity density at mid-rapidity in Pb-Pb collisions at $\sqrt{s_{NN}}=2.76$ TeV,''
  Phys.\ Rev.\ Lett.\  {\bf 106}, 032301 (2011).



\bibitem{Acharya:2018qsh} 
  S.~Acharya {\it et al.} [ALICE Collaboration],
  ``Transverse momentum spectra and nuclear modification factors of charged particles in pp, p-Pb and Pb-Pb collisions at the LHC,''
  JHEP {\bf 1811}, 013 (2018).



\bibitem{Pratt:2015zsa} 
  S.~Pratt, E.~Sangaline, P.~Sorensen and H.~Wang,
  ``Constraining the Eq. of State of Super-Hadronic Matter from Heavy-Ion Collisions,''
  Phys.\ Rev.\ Lett.\  {\bf 114}, 202301 (2015).



\bibitem{Adare:2013esx} 
  A.~Adare {\it et al.} [PHENIX Collaboration],
  ``Spectra and ratios of identified particles in Au+Au and $d$+Au collisions at $\sqrt{s_{NN}}=200$ GeV,''
  Phys.\ Rev.\ C {\bf 88}, no. 2, 024906 (2013).


\bibitem{Bjorken:1982qr} 
  J.~D.~Bjorken,
  ``Highly Relativistic Nucleus-Nucleus Collisions: The Central Rapidity Region,''
  Phys.\ Rev.\ D {\bf 27}, 140-151 (1983).



\bibitem{Kolb:2003dz} 
  P.~F.~Kolb and U.~W.~Heinz,
  ``Hydrodynamic description of ultrarelativistic heavy ion collisions,''
  In *Hwa, R.C. (ed.) et al.: Quark gluon plasma* 634-714
  [nucl-th/0305084].



\bibitem{Vredevoogd:2008id} 
  J.~Vredevoogd and S.~Pratt,
  ``Universal Flow in the First Stage of Relativistic Heavy Ion Collisions,''
  Phys.\ Rev.\ C {\bf 79}, 044915 (2009).



\bibitem{vanderSchee:2013pia} 
  W.~van der Schee, P.~Romatschke and S.~Pratt,
  ``Fully Dynamical Simulation of Central Nuclear Collisions,''
  Phys.\ Rev.\ Lett.\  {\bf 111}, no. 22, 222302 (2013).



\bibitem{Keegan:2016cpi} 
  L.~Keegan, A.~Kurkela, A.~Mazeliauskas and D.~Teaney,
  ``Initial conditions for hydrodynamics from weakly coupled pre-equilibrium evolution,''
  JHEP {\bf 1608}, 171 (2016).



\bibitem{Eskola:2000xq} 
  K.~J.~Eskola, K.~Kajantie and K.~Tuominen,
  ``Centrality dependence of multiplicities in ultrarelativistic nuclear collisions,''
  Phys.\ Lett.\ B {\bf 497}, 39-43 (2001).



\bibitem{Kolb:2001qz} 
  P.~F.~Kolb, U.~W.~Heinz, P.~Huovinen, K.~J.~Eskola and K.~Tuominen,
  ``Centrality dependence of multiplicity, transverse energy, and elliptic flow from hydrodynamics,''
  Nucl.\ Phys.\ A {\bf 696}, 197-215 (2001).



\bibitem{Miller:2007ri} 
  M.~L.~Miller, K.~Reygers, S.~J.~Sanders and P.~Steinberg,
  ``Glauber modeling in high energy nuclear collisions,''
  Ann.\ Rev.\ Nucl.\ Part.\ Sci.\  {\bf 57}, 205-243 (2007).



\bibitem{Abelev:2013qoq} 
  B.~Abelev {\it et al.} [ALICE Collaboration],
  ``Centrality determination of Pb-Pb collisions at $\sqrt{s_{NN}}$ = 2.76 TeV with ALICE,''
  Phys.\ Rev.\ C {\bf 88}, no. 4, 044909 (2013).



\bibitem{Moreland:2014oya} 
  J.~S.~Moreland, J.~E.~Bernhard and S.~A.~Bass,
  ``Alternative ansatz to wounded nucleon and binary collision scaling in high-energy nuclear collisions,''
  Phys.\ Rev.\ C {\bf 92}, no. 1, 011901 (2015).



\bibitem{Hama:2004rr} 
  Y.~Hama, T.~Kodama and O.~Socolowski, Jr.,
  ``Topics on hydrodynamic model of nucleus-nucleus collisions,''
  Braz.\ J.\ Phys.\  {\bf 35}, 24-51 (2005).



\bibitem{Broniowski:2009fm} 
  W.~Broniowski, M.~Chojnacki and L.~Obara,
  ``Size fluctuations of the initial source and the event-by-event transverse momentum fluctuations in relativistic heavy-ion collisions,''
  Phys.\ Rev.\ C {\bf 80}, 051902 (2009).



\bibitem{Schenke:2010nt} 
  B.~Schenke, S.~Jeon and C.~Gale,
  ``(3+1)D hydrodynamic simulation of relativistic heavy-ion collisions,''
  Phys.\ Rev.\ C {\bf 82}, 014903 (2010).



\bibitem{Schenke:2011bn} 
  B.~Schenke, S.~Jeon and C.~Gale,
  ``Higher flow harmonics from (3+1)D event-by-event viscous hydrodynamics,''
  Phys.\ Rev.\ C {\bf 85}, 024901 (2012).



\bibitem{Paquet:2015lta} 
  J.~F.~Paquet, C.~Shen, G.~S.~Denicol, M.~Luzum, B.~Schenke, S.~Jeon and C.~Gale,
  ``Production of photons in relativistic heavy-ion collisions,''
  Phys.\ Rev.\ C {\bf 93}, no. 4, 044906 (2016).



\bibitem{Heinz:2013th} 
  U.~Heinz and R.~Snellings,
  ``Collective flow and viscosity in relativistic heavy-ion collisions,''
  Ann.\ Rev.\ Nucl.\ Part.\ Sci.\  {\bf 63}, 123-151 (2013).



\bibitem{Cooper:1974mv} 
  F.~Cooper and G.~Frye,
  ``Comment on the Single Particle Distribution in the Hydrodynamic and Statistical Thermodynamic Models of Multiparticle Production,''
  Phys.\ Rev.\ D {\bf 10}, 186-189 (1974).


\bibitem{Bazavov:2018mes} 
  A.~Bazavov {\it et al.} [HotQCD Collaboration],
  ``Chiral crossover in QCD at zero and non-zero chemical potentials,''
  Phys.\ Lett.\ B {\bf 795}, 15-21 (2019).




\bibitem{Gale:2012rq} 
  C.~Gale, S.~Jeon, B.~Schenke, P.~Tribedy and R.~Venugopalan,
  ``Event-by-event anisotropic flow in heavy-ion collisions from combined Yang-Mills and viscous fluid dynamics,''
  Phys.\ Rev.\ Lett.\  {\bf 110}, no. 1, 012302 (2013).



\bibitem{Eskola:2017bup} 
  K.~J.~Eskola, H.~Niemi, R.~Paatelainen and K.~Tuominen,
  ``Predictions for multiplicities and flow harmonics in 5.44 TeV Xe+Xe collisions at the CERN Large Hadron Collider,''
  Phys.\ Rev.\ C {\bf 97}, no. 3, 034911 (2018).



\bibitem{Weller:2017tsr} 
  R.~D.~Weller and P.~Romatschke,
  ``One fluid to rule them all: viscous hydrodynamic description of event-by-event central p+p, p+Pb and Pb+Pb collisions at $\sqrt{s}=5.02$ TeV,''
  Phys.\ Lett.\ B {\bf 774}, 351-356 (2017).



\bibitem{Dubla:2018czx} 
  A.~Dubla, S.~Masciocchi, J.~M.~Pawlowski, B.~Schenke, C.~Shen and J.~Stachel,
  ``Towards QCD-assisted hydrodynamics for heavy-ion collision phenomenology,''
  Nucl.\ Phys.\ A {\bf 979}, 251-264 (2018).



\bibitem{Teaney:2003kp} 
  D.~Teaney,
  ``The Effects of viscosity on spectra, elliptic flow, and HBT radii,''
  Phys.\ Rev.\ C {\bf 68}, 034913 (2003).



\bibitem{Dusling:2009df} 
  K.~Dusling, G.~D.~Moore and D.~Teaney,
  ``Radiative energy loss and v(2) spectra for viscous hydrodynamics,''
  Phys.\ Rev.\ C {\bf 81}, 034907 (2010).



\bibitem{Bozek:2009dw} 
  P.~Bozek,
  ``Bulk and shear viscosities of matter created in relativistic heavy-ion collisions,''
  Phys.\ Rev.\ C {\bf 81}, 034909 (2010).



\bibitem{Teaney:2000cw} 
  D.~Teaney, J.~Lauret and E.~V.~Shuryak,
  ``Flow at the SPS and RHIC as a quark gluon plasma signature,''
  Phys.\ Rev.\ Lett.\  {\bf 86}, 4783-4786 (2001).



\bibitem{Petersen:2008dd} 
  H.~Petersen, J.~Steinheimer, G.~Burau, M.~Bleicher and H.~Stocker,
  ``A Fully Integrated Transport Approach to Heavy Ion Reactions with an Intermediate Hydrodynamic Stage,''
  Phys.\ Rev.\ C {\bf 78}, 044901 (2008).



\bibitem{Petersen:2009mz} 
  H.~Petersen, J.~Steinheimer, M.~Bleicher and H.~Stocker,
  ``<m(T)> excitation function: Freeze-out and equation of state dependence,''
  J.\ Phys.\ G {\bf 36}, 055104 (2009).



\bibitem{Luo:2015doi} 
  X.~Luo,
  ``Exploring the QCD Phase Structure with Beam Energy Scan in Heavy-ion Collisions,''
  Nucl.\ Phys.\ A {\bf 956}, 75-82 (2016).



\bibitem{Monnai:2017cbv} 
  A.~Monnai and J.~Y.~Ollitrault,
  ``Constraining the equation of state with identified particle spectra,''
  Phys.\ Rev.\ C {\bf 96}, no. 4, 044902 (2017),

\end{thebibliography}
\end{document}